# Twisted monolayer and bilayer graphene for vertical tunneling transistors

Davit A. Ghazaryan [1, 2], Abhishek Misra[3, 4], Evgenii E. Vdovin[5], Kenji Watanabe[6], Takashi Taniguchi[6], Sergei V. Morozov[5], Artem Mishchenko[1, 3], Kostya S. Novoselov [1, 3, 7, 8,*]

[1]*Department of Physics and Astronomy, University of Manchester, Manchester, M13 9PL, UK.*

[2]*Center for Photonics and 2D Materials, Moscow Institute of Physics and Technology, Dolgoprudny, 141701, Russian Federation.*

[3]*National Graphene Institute, University of Manchester, Manchester, M13 9PL, UK.*

[4]*Department of Physics, Indian Institute of Technology Madras (IIT Madras), Chennai, 600036, India.*

[5]*Institute for Problems of Microelectronics Technologies and High-Purity Materials, Russian Academy of Sciences, Chernogolovka, 142432, Russian Federation.*

[6]*National Institute for Materials Science, Namiki 1-1, Tsukuba, 305-0044, Ibaraki, Japan.*

[7]*Centre for Advanced 2D Materials, National University of Singapore, 1147546, Singapore.*

[8]*Chongqing 2D Materials Institute, Chongqing, 400707, China.*

[*]*Authors to whom correspondence should be addressed:* kostya@manchester.ac.uk

**We prepare twist-controlled resonant tunneling transistors consisting of monolayer (Gr) and Bernal bilayer (BGr) graphene electrodes separated by a thin layer of hexagonal boron nitride (hBN). The resonant conditions are achieved by closely aligning the crystallographic orientation of the graphene electrodes, which leads to momentum conservation for tunneling electrons at certain bias voltages. Under such conditions, negative differential conductance (NDC) can be achieved. Application of in-plane magnetic field leads to electrons acquiring additional momentum during the tunneling process, which allows control over the resonant conditions.**

Gate-controlled vertical tunneling transistors[1–5] are of significant interest for studying intrinsic properties of crystalline two-dimensional (2D) materials forming either electrodes[6–10] or the tunnel barrier[11–15]. In particular, the use of hexagonal boron nitride (hBN) as a tunnel barrier, incorporated between graphene (Gr) electrodes allows formation of high-quality tunneling transistors with non-trivial properties[16,17]. Such devices benefit from establishment of atomically smooth interfaces[18], which suppresses any momentum scattering during tunneling, making possible observation of different types of elastic and inelastic resonant effects. These include tunneling assisted by phonon emission[11,19,20], sequential tunneling through localized defect states[19,21], interstate percolation[22], and single-electron charging effects[23,24]. Most interestingly for this work – alignment of the crystallographic orientation of Gr electrodes[25–28] leads to observation of a number of resonances associated with the conservation of momentum, energy, and sublattice composition of tunneling electrons.

In this work, we study tunneling properties of devices formed by sandwiching hBN barrier between crystallographically aligned electrodes of Gr and Bernal bilayer graphene (BGr). Such devices demonstrate numerous resonances that are associated with momentum conservation conditions fulfilment when occupied (empty) states in Gr cross in momentum space with empty (occupied) states in BGr.

Our van der Waals heterostructures were formed by dry-transfer method[29,30] of high-quality, mechanically exfoliated graphene and hBN crystals. The whole heterostructure was built on a $Si^{++}/SiO_2$ wafer to allow gating and encapsulated with thick hBN (thhBN) for stability, accordingly, the complete device has a structure of $Si^{++}/SiO_2$/thhBN/*Gr*/*h*BN/*BGr*/thhBN. Electrodes to Gr and BGr were formed by e-beam lithography followed by deposition of Ti/Au contacts by e-beam evaporation.

Fig. 1(a and b) present a schematic illustration of crystallographically aligned Gr and BGr electrode layers along with the corresponding low-energy electronic bands at six corners of 1st Brillouin zone displaced by finite momentum $|\Delta K_\theta|$ due to the relative rotation of the crystals. Here, the momentum displacement is determined by a single parameter – the misalignment angle $\theta$, and can be expressed as

$$|\Delta K_\theta| = 4\pi\theta/3a ,\qquad(1)$$

where $a$ = 2.46 Å is graphene lattice constant.

To measure simultaneously *I-V* and differential tunneling conductance ($G$=d$I$/d$V$) characteristics, a small low-frequency AC excitation voltage was admixed to DC bias voltage $V$. At low bias voltage, no occupied states in either Gr or BGr cross with empty states in the opposite contact, resulting in low tunneling conductance. Applying finite bias, we simultaneously shift the bands in the two electrodes with respect to each other, and change the position of their Fermi levels. At certain biases, this results in occupied states in Gr or BGr (depending on the polarity of the bias voltage) to cross with the empty states in the opposite contact (Fig. 1(d)), resulting in the opening of a tunneling channel and increase in current and tunneling conductance, Fig. 1(e).

Fig. 2(a) presents the resulting *T* = 2 K contour map as a function of bias and gate voltages. A small asymmetry of the picture concerning zero gate voltage is attributed to small unintentional doping of the electrodes. As expected, at low bias and gate voltages the tunneling conductance is suppressed. However, several characteristic resonances are observed at finite bias and gate voltages.

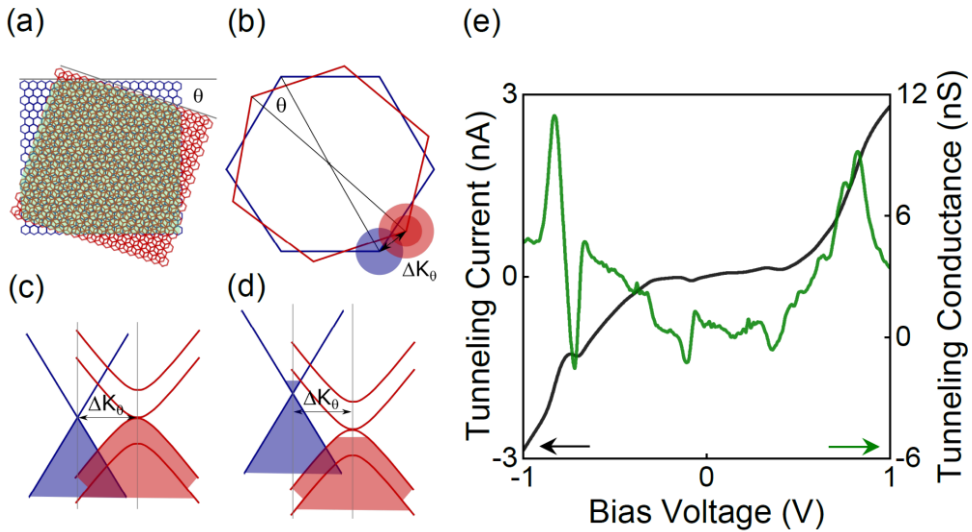

Fig. 1. Initial characterization of vertical tunneling transistors with twisted Gr and BGr electrodes. (a) Schematic illustration of electrodes introducing crystal lattice misalignment angle $\theta$, and (b) corresponding $\Delta K_\theta$ displacement of electronic bands. For simplicity Fermi surfaces are shaped at single corner of the 1st Brillouin zone. (c) Diagrammatic band structure for the case of zero bias voltage. (d) Same as (c), but for finite bias voltage. (e) Typical zero-gate current/conductance-bias voltage characteristics measured at *T* = 2 K.

To analyze the tunneling resonances observed in our devices, we construct a self-consistent electrostatic model using the parallel-plate capacitor model (and following procedures described in the previous works[8,25,26,28]). Accounting for a partial Gr screening of electric field generated from the gate electrode and also for internal self-screening of BGr bandgap, we find the dependences of the chemical potential in Gr ($\mu_{Gr}$) and BGr ($\mu_{BGr}$), energy band offset $\Delta\varphi$, and electrically tunable BGr bandgap $\Delta_g$ on gate and bias voltages. We then evaluate the resonant conditions (taken as an event when any of the occupied states in one electrode cross in momentum and energy space with empty states in the other electrode) for a given angle of crystallographic alignment between Gr and BGr crystal lattices. In our model, the low-energy electronic band structure of BGr[31,32] is expressed by

$$\varepsilon_{m,n}(k,\Delta_g) = m\sqrt{\gamma_1^2/2 + \Delta_g^2/4 + \hbar^2 v_F^2 k^2 + n\sqrt{\gamma_1^4/4 + \hbar^2 v_F^2 k^2(\gamma_1^2 + \Delta_g^2)}}, \quad (2)$$

where $m = \pm 1$ and $n = \mp 1$ stand for the description of 1st conduction and valence bands, $m = \pm 1$ and $n = \pm 1$ for 2nd bands, $\hbar$ is the reduced Planck's constant, $v_F$ is Fermi velocity, and $\gamma_1$ is the interlayer coupling parameter (see Supplementary Material for further details).

Fig. 2(b) presents resonant conditions for several elastic (Fig. 2(c)) and inelastic (Fig. 2(d)) events depending on the gate and bias voltages. It fits those observed within the experiment (Fig. 2(a)) very well. Note, that the only fitting parameter here is the twist angle between Gr and BGr, which was found to be $\theta = 3.2°$. The grey dash-dot line represents the conditions when the Fermi level in Gr being exactly at the Dirac point. Its linear dependence in the gate-bias voltage plane represents the fact that the density of states in BGr is constant (it would be square root-like if both electrodes are Gr[28] because of the linear DoS). Fitting of the slope of this line allows determining the thickness of the hBN tunnel barrier – which was found to be 5 layers thick in our case.

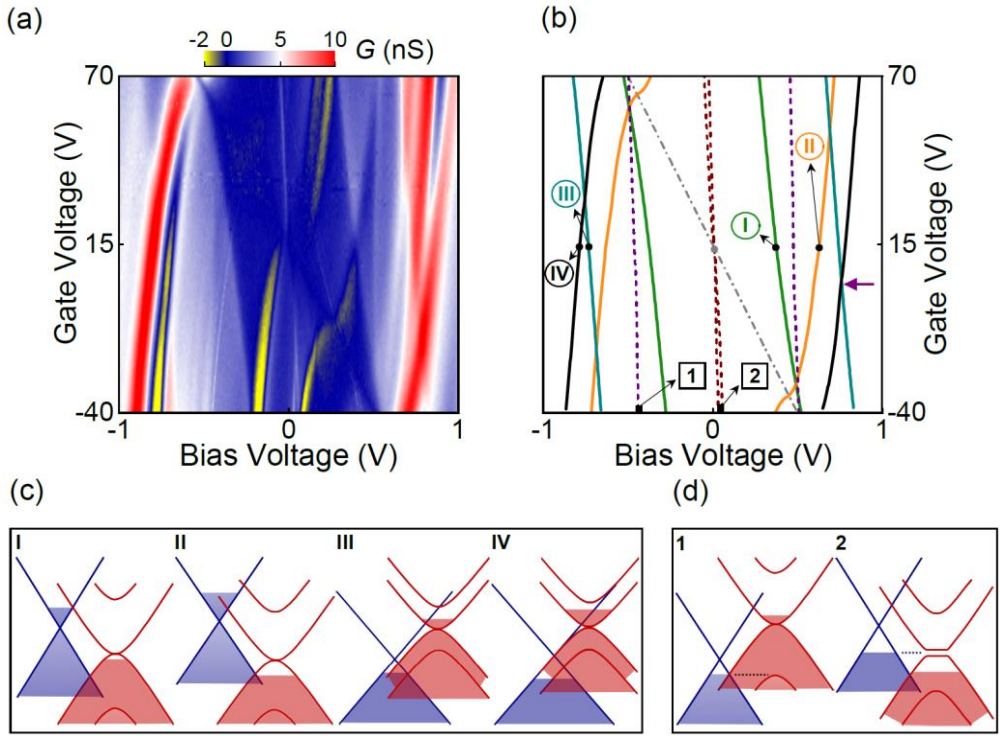

Fig. 2. The typical pattern of conductive channels of vertical tunneling transistors with twisted Gr and BGr electrodes. (a) $T$ = 2 K differential tunneling conductance dependence on gate and bias voltages. (b) The characteristic pattern of DoS lines indicating the channels for resonant tunneling transition at $\theta \approx 3.2°$. Numerated black dotes denote conductive channels depicted in (c) and corresponding to elastic transitions. Rectangles denote inelastic conductive channels depicted in (d). (c) Diagrams of the band structures visualizing band alignments and potential distributions for elastic resonant events. (d) Same as (c), but for inelastic events.

To illustrate the resonant tunneling transitions associated with $|\Delta K_\theta|$ momentum displacement, we introduce schematic band diagrams for the different resonant conditions identified in our devices, Fig. 2(c). The notable tunneling channels get activated when the chemical potential of Gr aligns with the crossing point between the Gr cone and the 1$^{st}$ conduction (conditions I) or valence bands in BGr. These events are represented by two lines in Fig. 2(b) and correspond to step-like resonances in differential tunneling conductance. Thus, the set of green lines introduce Gr to BGr tunneling for the condition of $k \rightarrow (\mu_{Gr}/\hbar v_F) \pm \Delta K_\theta$ at $\mu_{Gr} + \Delta\varphi = \varepsilon_{\pm,\mp}(k, \Delta_g)$.

The set of orange lines introduce the opposite tunneling direction for the conditions of $k \rightarrow ((\pm\Delta\varphi \mp \mu_{BGr})/\hbar v_F) + \Delta K_\theta$ at $\mu_{BGr} = \varepsilon_{\pm,\mp}(k, \Delta_g)$, when the chemical potential in BGr reaches the crossing points between Gr and BGr bands (condition II in Fig. 2(c)). Another set of step resonance channels activate when the chemical potential of Gr reaches the crossing point between the Gr cone and 2$^{nd}$ bands of BGr. The characteristic blue lines in Fig. 2(b) present the position of the resonance, and Fig. 2 (c, III) displays the band diagrams at negative bias voltages. In this case, the resonances emerge when $k \rightarrow (\mu_{Gr}/\hbar v_F) \pm \Delta K_\theta$ at $\mu_{Gr} + \Delta\varphi = \varepsilon_{\pm,\pm}(k, \Delta_g)$.

There is also one resonance, which is associated with the alignment of the bands themselves (rather than with the alignment of a Fermi level with a particular band). Since the slope of the Dirac cones in Gr matches that in BGr at high energies, one can achieve a situation when a substantial part of the Dirac cone become nested to the bands of BGr, leading to resonant conditions being fulfilled for many electronic states simultaneously. Such resonances are schematically presented in Fig. 2(c, IV), depicted by black lines in Fig. 2(b), and correspond to the condition when $k \rightarrow \Delta K_\theta$ at $\Delta\varphi = \varepsilon_{\pm,\mp}(k, \Delta_g)$. For this type of resonance, all the states either simultaneously participate in the tunneling or become off-resonance. This leads to the appearance of the NDC region, see Fig. 1(e) at negative bias voltage. Note, that the NDC observed in Gr-BGr tunneling devices are substantially weaker than similar resonances observed for tunneling between two Gr electrodes[28], as in the latter case the nesting is exact.

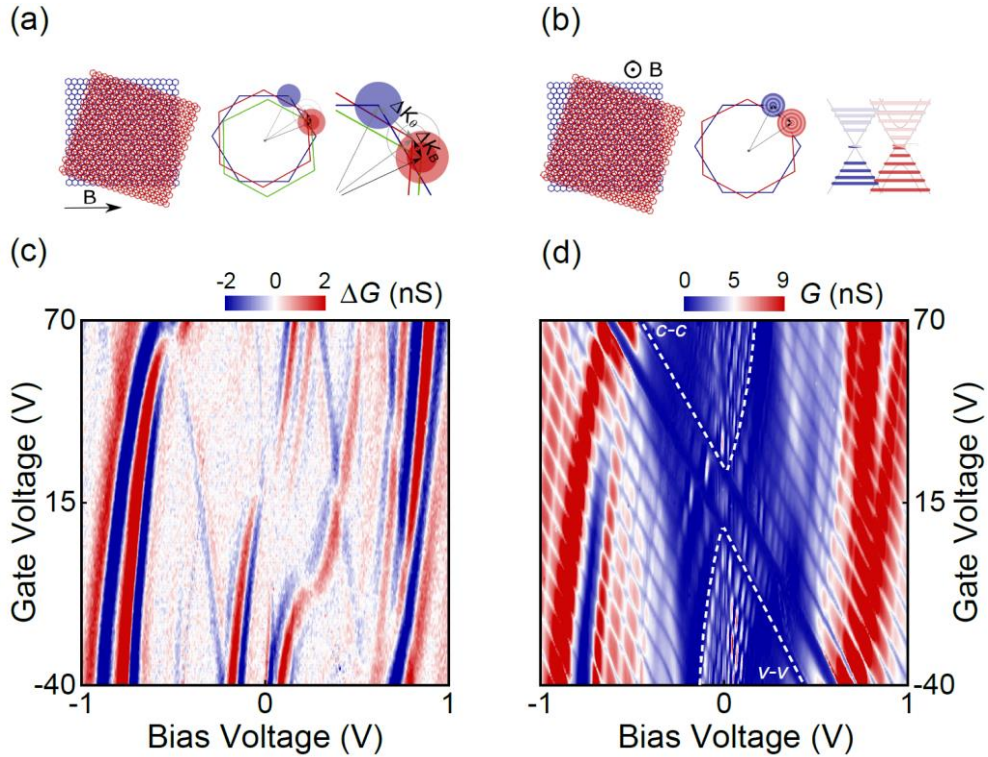

Fig. 3. *T* = 2K magnetotunneling transitions in vertical tunneling transistors with twisted Gr and BGr electrodes. (a) Schematic illustration of effective displacement of electronic bands of collector electrode arisen due to subjection to strong parallel magnetic fields. (b) Schematic illustration of the quantized spectrum of Landau Levels arisen due to subjection to perpendicular magnetic fields. (c). $B_\parallel$ = 15 T to zero-field difference contour map of differential tunneling conductance measured as a function of bias and gate voltages. (d) $B_\perp$ = 5 T contour map of differential tunneling conductance measured as a function of bias and gate voltages. Dashed lines highlight regions of tunneling transitions in equivalent bands.

We also observe weaker resonances that are associated with inelastic processes, which are not sensitive to the twist angle. The reason such resonances are observed (even though weaker than the elastic ones) is the enhanced DoS at the van Hove singularities[31,32] of BGr. For instance, the event depicted in Fig. 2(e, 1) corresponds to the purple lines in Fig. 2(b). Here, the resonances emerge when the chemical potential of Gr aligns to the bottom (top) edge of the 2$^{nd}$ conductance (valence) band of BGr, $\mu_{Gr} + \Delta\varphi = \varepsilon_{\pm,\pm}(k, \Delta_g)$. For the other case (Fig. 2(e, 2)), described by brown lines in Fig. 2(b), the resonances emerge when the chemical potential of Gr aligns with the edges of BGr bandgap, $\mu_{Gr} + \Delta\varphi = \varepsilon_{\pm,\mp}(0, \Delta_g)$.

In-plane (perpendicular to the tunneling current) magnetic field can shift the position of the resonances. It is easy to understand this in semiclassical interpretation as electrons acquiring additional momentum during tunneling process due to the Lorentz force[26,28]. The additional momentum acquired in the magnetic field can be expressed as

$$|\Delta K_B| = \frac{edB_{\parallel}}{\hbar}, \tag{3}$$

where $e$ is the elementary charge and $d$ is the tunnel barrier thickness. However, unlike $|\Delta K_\theta|$, which is a result of the rotation of the two Brillouin zones, the additional momentum acquired in the magnetic field results in parallel shifts of the Brillouin zones of Gr and BGr with respect to each other. Thus, the overall influence of magnetic field and rotation will result in different momentum displacement for different corners of the Brillouin zone, Fig. 3(a), and thus, in the change of resonant conditions.

Because of the small thickness of the tunnel barrier, the $|\Delta K_B|$ is very small. The best way to demonstrate the shift in the resonance positions in the magnetic field is to plot the difference of tunneling conductance with and without magnetic field: $\Delta G(B_0)=G(B_{II}=B_0)-G(B_{II}=0)$. Measuring such changes in the differential tunneling conductance at strong parallel magnetic fields of $B_{\parallel}$ = 15 T, we observe a finite voltage shift for the conductive channels arising from elastic resonant tunneling transitions, Fig. 3(c). Note, that the resonances associated with inelastic transitions are not visible on this map, as they do not depend on the momentum shift. A small signal in the region of the resonance associated with tunneling into the van Hove singularities at the BGr bandgap (Fig. 2 (c, 2), brown line in Fig. 2(b)) is associated with bandgap modification due to interaction-induced and Zeeman effects.

On the other hand, when presenting magnetic fields perpendicular to electrodes, one expects the emergence of cyclotron gaps, and hence, a quantization of electronic bands[33] of the graphene electrodes, Fig. 3(b). This results in the formation of additional sets of resonances due to the quantization of electronic bands to the Landau Levels. Notably, for Gr (BGr), the LLs are distributed nonequidistantly[34] (equidistantly) in energy. In $B_\perp$ = 5 T magnetic fields, measuring differential tunneling conductance contour reveals such resonances, Fig. 3(d). These follow the main electrostatic pattern of Gr-BGr and correspond to inter LL resonant tunneling transitions. Furthermore, we observe an enhancement of such resonances generated by tunneling transitions between conduction - conduction or valence - valence electronic bands (in contrast to conduction-valence or valence-conduction). This behavior is attributed to the manifestation of sub-lattice (chirality) composition of graphene electrons. Within the regions of equivalent bands, depicted in Fig. 3(d) by dashed lines, the conservation of chirality of tunneling electrons, escalates the transition rate[27,35].

In summary, we demonstrated that twist-controlled resonant tunneling transistors based on Gr and BGr electrodes demonstrate a set of gate- and bias-tuneable resonances. These resonances emerge owing to resonant tunneling transitions of electrons that conserve in-plane momentum, energy, and chirality. Such tunneling transitions can result in the NDC region. Strong parallel magnetic fields can shift their position, whereas perpendicular magnetic fields can activate extra resonant features. The latter appear amplified for the areas of tunneling transitions within equivalent electronic bands.

**Supplementary Material**

See the Supplementary Material for devices with other crystal lattice misalignment angles of Gr-BGr, notes on electrostatic model and fabrication procedure.

**Acknowledgements**

We would like to thank John Wallbank for useful discussions. This work was supported by EU Flagship Programs (Graphene CNECTICT-604391 and 2D-SIPC Quantum Technology), European Research Council Synergy Grant Hetero2D, the Royal Society, EPSRC grants EP/N010345/1, EP/P026850/1, EP/S030719/1. E. E. E. Vdovin and S. V. Morozov acknowledge the support from the Russian Science Foundation (17–12–01393).

**References**

[1] L. Britnell, R. V. Gorbachev, R. Jalil, B.D. Belle, F. Schedin, A. Mishchenko, T. Georgiou, M.I. Katsnelson, L. Eaves, S. V. Morozov, N.M.R. Peres, J. Leist, A.K. Geim, K.S. Novoselov, and L.A. Ponomarenko, Science **335**, 947 (2012).

[2] L. Britnell, R. V. Gorbachev, R. Jalil, B.D. Belle, F. Schedin, M.I. Katsnelson, L. Eaves, S. V. Morozov, A.S. Mayorov, N.M.R. Peres, A.H. Castro Neto, J. Leist, A.K. Geim, L.A. Ponomarenko, and K.S. Novoselov, Nano Letters **12**, 1707 (2012).

[3] L. Britnell, R. V. Gorbachev, A.K. Geim, L.A. Ponomarenko, A. Mishchenko, M.T. Greenaway, T.M. Fromhold, K.S. Novoselov, and L. Eaves, Nature Communications **4**, (2013).

[4] G.H. Lee, Y.J. Yu, C. Lee, C. Dean, K.L. Shepard, P. Kim, and J. Hone, Applied Physics Letters **99**, (2011).

[5] Y. Lv, W. Qin, C. Wang, L. Liao, and X. Liu, Advanced Electronic Materials **5**, (2019).

[6] T. Roy, M. Tosun, X. Cao, H. Fang, D.H. Lien, P. Zhao, Y.Z. Chen, Y.L. Chueh, J. Guo, and A. Javey, ACS Nano **9**, 2071 (2015).

[7] B. Fallahazad, K. Lee, S. Kang, J. Xue, S. Larentis, C. Corbet, K. Kim, H.C.P. Movva, T. Taniguchi, K. Watanabe, L.F. Register, S.K. Banerjee, and E. Tutuc, Nano Letters **15**, 428 (2015).

[8] M. Zhu, D. Ghazaryan, S.K. Son, C.R. Woods, A. Misra, L. He, T. Taniguchi, K. Watanabe, K.S. Novoselov, Y. Cao, and A. Mishchenko, 2D Materials **4**, (2017).

[9] R. Cheng, F. Wang, L. Yin, K. Xu, T. Ahmed Shifa, Y. Wen, X. Zhan, J. Li, C. Jiang, Z. Wang, and J. He, Applied Physics Letters **110**, (2017).

[10] G.W. Burg, N. Prasad, K. Kim, T. Taniguchi, K. Watanabe, A.H. Macdonald, L.F. Register, and E. Tutuc, Physical Review Letters **120**, (2018).

[11] S. Jung, M. Park, J. Park, T.Y. Jeong, H.J. Kim, K. Watanabe, T. Taniguchi, D.H. Ha, C. Hwang, and Y.S. Kim, Scientific Reports **5**, (2015).

[12] D. Ghazaryan, M.T. Greenaway, Z. Wang, V.H. Guarochico-Moreira, I.J. Vera-Marun, J. Yin, Y. Liao, S. V. Morozov, O. Kristanovski, A.I. Lichtenstein, M.I. Katsnelson, F. Withers, A. Mishchenko, L. Eaves, A.K. Geim, K.S. Novoselov, and A. Misra, Nature Electronics **1**, 344 (2018).

[13] D.R. Klein, D. MacNeill, J.L. Lado, D. Soriano, E. Navarro-Moratalla, K. Watanabe, T. Taniguchi, S. Manni, P. Canfield, J. Fernández-Rossier, and P. Jarillo-Herrero, Science **360**, 1218 (2018).

[14] X. Jiang, X. Shi, M. Zhang, Y. Wang, Z. Gu, L. Chen, H. Zhu, K. Zhang, Q. Sun, and D.W. Zhang, ACS Applied Nano Materials **2**, 5674 (2019).

[15] J. Kang, D. Jariwala, C.R. Ryder, S.A. Wells, Y. Choi, E. Hwang, J.H. Cho, T.J. Marks, and M.C. Hersam, Nano Letters **16**, 2580 (2016).

[16] Y. Liu, N.O. Weiss, X. Duan, H.C. Cheng, Y. Huang, and X. Duan, Nature Reviews Materials **1**, (2016).

[17] K.S. Novoselov, A. Mishchenko, A. Carvalho, and A.H. Castro Neto, Science **353**, (2016).

[18] C.R. Dean, A.F. Young, I. Meric, C. Lee, L. Wang, S. Sorgenfrei, K. Watanabe, T. Taniguchi, P. Kim, K.L. Shepard, and J. Hone, Nature Nanotechnology **5**, 722 (2010).

[19] U. Chandni, K. Watanabe, T. Taniguchi, and J.P. Eisenstein, Nano Letters **16**, 7982 (2016).

[20] E.E. Vdovin, A. Mishchenko, M.T. Greenaway, M.J. Zhu, D. Ghazaryan, A. Misra, Y. Cao, S. V. Morozov, O. Makarovsky, T.M. Fromhold, A. Patanè, G.J. Slotman, M.I. Katsnelson, A.K. Geim, K.S. Novoselov, and L. Eaves, Physical Review Letters **116**, (2016).

[21] U. Chandni, K. Watanabe, T. Taniguchi, and J.P. Eisenstein, Nano Letters **15**, 7329 (2015).

[22] M.T. Greenaway, E.E. Vdovin, D. Ghazaryan, A. Misra, A. Mishchenko, Y. Cao, Z. Wang, J.R. Wallbank, M. Holwill, Y.N. Khanin, S. V. Morozov, K. Watanabe, T. Taniguchi, O. Makarovsky, T.M. Fromhold, A. Patanè, A.K. Geim, V.I. Fal'ko, K.S. Novoselov, and L. Eaves, Communications Physics **1**, (2018).

[23] R.M. Feenstra, D. Jena, and G. Gu, Journal of Applied Physics **111**, (2012).

[24] G. Kim, S.S. Kim, J. Jeon, S.I. Yoon, S. Hong, Y.J. Cho, A. Misra, S. Ozdemir, J. Yin, D. Ghazaryan, M. Holwill, A. Mishchenko, D. V. Andreeva, Y.J. Kim, H.Y. Jeong, A.R. Jang, H.J. Chung, A.K. Geim, K.S. Novoselov, B.H. Sohn, and H.S. Shin, Nature Communications **10**, (2019).

[25] T.L.M. Lane, J.R. Wallbank, and V.I. Fal'ko, Applied Physics Letters **107**, (2015).

[26] J.R. Wallbank, D. Ghazaryan, A. Misra, Y. Cao, J.S. Tu, B.A. Piot, M. Potemski, S. Pezzini, S. Wiedmann, U. Zeitler, T.L.M. Lane, S. V. Morozov, M.T. Greenaway, L. Eaves, A.K. Geim, V.I. Fal'ko, K.S. Novoselov, and A. Mishchenko, Science **353**, 575 (2016).

[27] M.T. Greenaway, E.E. Vdovin, A. Mishchenko, O. Makarovsky, A. Patanè, J.R. Wallbank, Y. Cao, A. V. Kretinin, M.J. Zhu, S. V. Morozov, V.I. Fal'Ko, K.S. Novoselov, A.K. Geim, T.M. Fromhold, and L. Eaves, Nature Physics **11**, 1057 (2015).

[28] A. Mishchenko, J.S. Tu, Y. Cao, R. V. Gorbachev, J.R. Wallbank, M.T. Greenaway, V.E. Morozov, S. V. Morozov, M.J. Zhu, S.L. Wong, F. Withers, C.R. Woods, Y.J. Kim, K. Watanabe, T. Taniguchi, E.E. Vdovin, O. Makarovsky, T.M. Fromhold, V.I. Fal'ko, A.K. Geim, L. Eaves, and K.S. Novoselov, Nature Nanotechnology **9**, 808 (2014).

[29] L. Wang, I. Meric, P.Y. Huang, Q. Gao, Y. Gao, H. Tran, T. Taniguchi, K. Watanabe, L.M. Campos, D.A. Muller, J. Guo, P. Kim, J. Hone, K.L. Shepard, and C.R. Dean, Science **342**, 614 (2013).

[30] A. V. Kretinin, Y. Cao, J.S. Tu, G.L. Yu, R. Jalil, K.S. Novoselov, S.J. Haigh, A. Gholinia, A. Mishchenko, M. Lozada, T. Georgiou, C.R. Woods, F. Withers, P. Blake, G. Eda, A. Wirsig, C. Hucho, K. Watanabe, T. Taniguchi, A.K. Geim, and R. V. Gorbachev, Nano Letters **14**, 3270 (2014).

[31] E. McCann, D.S.L. Abergel, and V.I. Fal'ko, European Physical Journal: Special Topics **148**, 91 (2007).

[32] E. McCann and M. Koshino, Reports on Progress in Physics **76**, (2013).

[33] K.S. Novoselov, A.K. Geim, S. V. Morozov, D. Jiang, M.I. Katsnelson, I. V. Grigorieva, S. V. Dubonos, and A.A. Firsov, Nature **438**, 197 (2005).

[34] A.H. Castro Neto, F. Guinea, N.M.R. Peres, K.S. Novoselov, and A.K. Geim, Reviews of Modern Physics **81**, 109 (2009).

[35] L. Pratley and U. Zülicke, Physical Review B - Condensed Matter and Materials Physics **88**, (2013).

# Supplementary Material *for*
# Twisted monolayer and bilayer graphene for vertical tunneling transistors

*Davit A. Ghazaryan*[1,2], *Abhishek Misra*[3,4], *Evgenii E. Vdovin*[5], *Kenji Watanabe*[6], *Takashi Taniguchi*[6], *Sergei V. Morozov*[5], *Artem Mishchenko*[1,3], *Kostya S. Novoselov*[1,3,7,8,*]

[1]Department of Physics and Astronomy, University of Manchester, Manchester, M13 9PL, UK.
[2]Center for Photonics and 2D Materials, Moscow Institute of Physics and Technology, Dolgoprudny, 141701, Russian Federation.
[3]National Graphene Institute, University of Manchester, Manchester, M13 9PL, UK.
[4]Department of Physics, Indian Institute of Technology Madras (IIT Madras), Chennai, 600036, India.
[5]Institute for Problems of Microelectronics Technologies and High-Purity Materials, Russian Academy of Sciences, Chernogolovka, 142432, Russian Federation.
[6]National Institute for Materials Science, Namiki 1-1, Tsukuba, 305-0044, Ibaraki, Japan.
[7]Centre for Advanced 2D Materials, National University of Singapore, 1147546, Singapore.
[8]Chongqing 2D Materials Institute, Chongqing, 400707, China.
[*]Authors to whom correspondence should be addressed: kostya@manchester.ac.uk

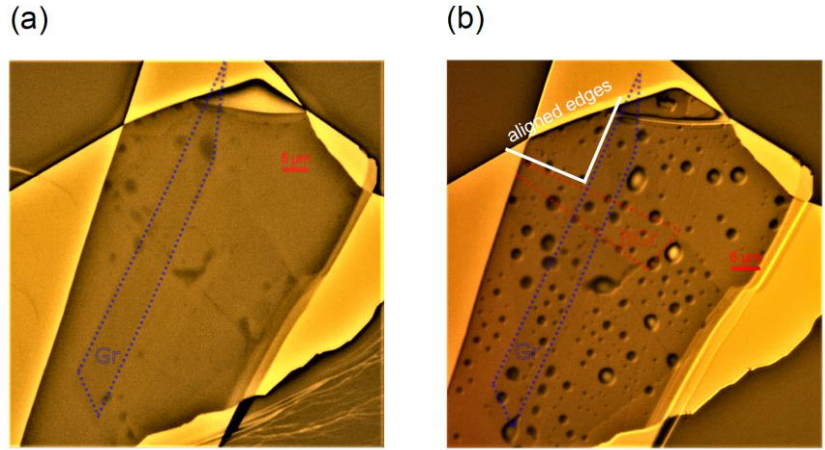

Fig. S1. Heterostructure preparation. Optical micrographs (a) after the assembly of thhBN-*Gr*/hBN layers, and (b) after the incorporation of /*BGr*-thhBN layers. Note the aligned crystallographic orientations of Gr and BGr electrodes achieved by choosing graphene strips with the edges oriented along the crystallographic directions, similar to[1]. Note the polymer residues at mid-interface (connecting hBN to BGr) of the heterostructure, and ~5 μm² bubble located at the cross-sectional area.

The simultaneous electrostatic equations for $Si^{++}/SiO_2$/thhBN-*Gr*/hBN/*BGr*-thhBN can be evaluated by a similar procedure as described in[2]. When accounting for the authentic band structure (eq. 2 of the main text) of BGr, one ends up with specific modifications to the relation between carrier density and chemical potential. It can be expressed by

$$n_{BGr}(\mu_{BGr}, \Delta_g) = \frac{\mu_{BGr}^2}{\pi \hbar^2 v_F^2} + \frac{\sqrt{\gamma_1^2 + \Delta_g^2}}{\pi \hbar^2 v_F^2}\left|\sqrt{\mu_{BGr}^2 - \frac{\gamma_1^2 \Delta_g^2}{4(\gamma_1^2+\Delta_g^2)}} - \sqrt{-\frac{\gamma_1^2 \Delta_g^2}{4(\gamma_1^2+\Delta_g^2)}}\right|, \quad (S1)$$

where $\hbar$ is the reduced Planck's constant, $v_F = 10^6$ m/s is the Fermi velocity, and $\gamma_1 = 0.39$ eV is the strongest interlayer coupling parameter.

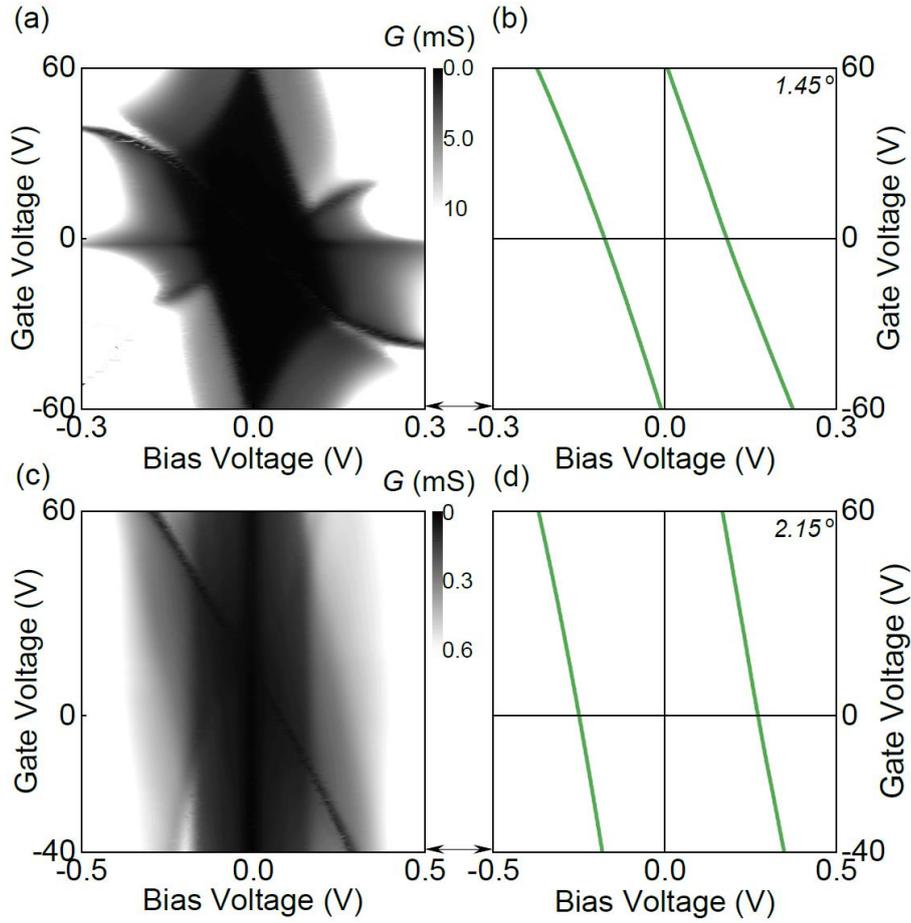

Fig. S2. Vertical tunneling transistors with different twists of Gr and BGr electrodes. (a) $T = 2$ K differential tunneling conductance dependence on gate and bias voltages for the device with monolayer hBN tunnel barrier. (b) The characteristic DoS lines indicating elastic tunneling channels of resonant condition I at $\theta \approx 1.45°$, and for monolayer hBN. (c) Same as (a), but for the device with trilayer hBN tunnel barrier. (d) The characteristic DoS lines indicating elastic tunneling channels of resonant condition I at $\theta \approx 2.15°$, and for trilayer hBN. Note the distortion of electrostatic pattern due to small resistances[3] (comparable to lateral resistances of graphene electrodes) of thin hBN barriers.

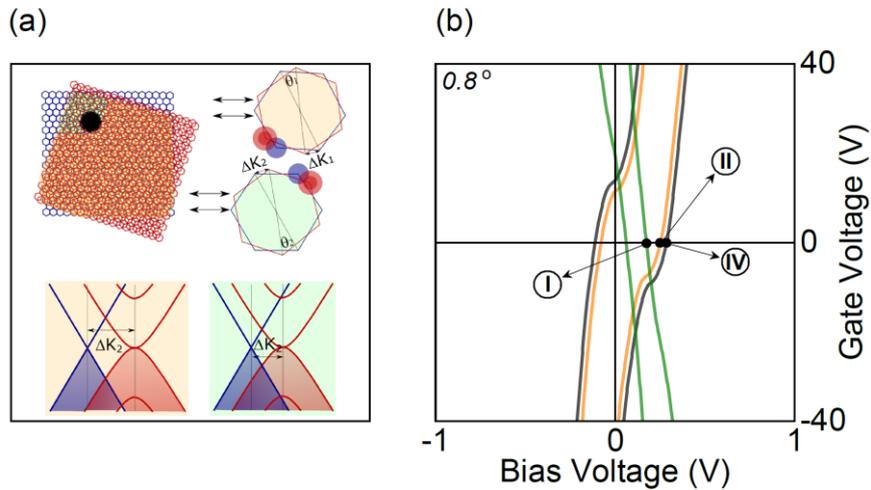

Fig. S3. Reconstruction of tunneling area. (a) Schematic illustration visualizing the reorganization of tunneling area[4] due to the mid-interfacial (hBN to BGr) bubbles. (b) Characteristic DoS lines for elastic tunneling transitions at $\theta_2 \approx 0.8°$ describing additional channels observed in Fig. 2(a). The sets of green, orange, and black lines introduce resonant conditions I, II, and IV, respectively.

The precise misalignment of Gr and BGr crystal lattices can be derived from the bias and gate voltage position of DoS lines of elastic resonant tunneling transitions, Fig. S2. For instance, from the resonant condition I, it can be expressed by

$$\theta = \frac{\sqrt{(\mu_{Gr}+\Delta\varphi)^2 + \Delta_g^2/4 - \sqrt{(\mu_{Gr}+\Delta\varphi)^2(\Delta_g^2+\gamma_1^2) - \gamma_1^2\Delta_g^2/4} \pm \mu_{Gr}}}{4\pi\hbar v_F/3a}, \qquad (S2)$$

where $a$ = 2.46 Å is graphene lattice constant.

**References**

[1] A. Mishchenko, J.S. Tu, Y. Cao, R. V. Gorbachev, J.R. Wallbank, M.T. Greenaway, V.E. Morozov, S. V. Morozov, M.J. Zhu, S.L. Wong, F. Withers, C.R. Woods, Y.J. Kim, K. Watanabe, T. Taniguchi, E.E. Vdovin, O. Makarovsky, T.M. Fromhold, V.I. Fal'ko, A.K. Geim, L. Eaves, and K.S. Novoselov, Nature Nanotechnology **9**, 808 (2014).

[2] M. Zhu, D. Ghazaryan, S.K. Son, C.R. Woods, A. Misra, L. He, T. Taniguchi, K. Watanabe, K.S. Novoselov, Y. Cao, and A. Mishchenko, 2D Materials **4**, (2017).

[3] L. Britnell, R. V. Gorbachev, R. Jalil, B.D. Belle, F. Schedin, M.I. Katsnelson, L. Eaves, S. V. Morozov, A.S. Mayorov, N.M.R. Peres, A.H. Castro Neto, J. Leist, A.K. Geim, L.A. Ponomarenko, and K.S. Novoselov, Nano Letters **12**, 1707 (2012).

[4] A. Weston, Y. Zou, V. Enaldiev, A. Summerfield, N. Clark, V. Zólyomi, A. Graham, C. Yelgel, S. Magorrian, M. Zhou, J. Zultak, D. Hopkinson, A. Barinov, T.H. Bointon, A. Kretinin, N.R. Wilson, P.H. Beton, V.I. Fal'ko, S.J. Haigh, and R. Gorbachev, Nature Nanotechnology **15**, 592 (2020).